\documentclass[twocolumn,english]{revtex4}
\usepackage[T1]{fontenc}
\usepackage[utf8]{inputenc}
\usepackage{color}
\usepackage{mathrsfs}
\usepackage{amsmath}
\usepackage{amssymb}
\usepackage{graphicx}
\usepackage{esint}

\makeatletter
\@ifundefined{textcolor}{}
{%
 \definecolor{BLACK}{gray}{0}
 \definecolor{WHITE}{gray}{1}
 \definecolor{RED}{rgb}{1,0,0}
 \definecolor{GREEN}{rgb}{0,1,0}
 \definecolor{BLUE}{rgb}{0,0,1}
 \definecolor{CYAN}{cmyk}{1,0,0,0}
 \definecolor{MAGENTA}{cmyk}{0,1,0,0}
 \definecolor{YELLOW}{cmyk}{0,0,1,0}
}

\@ifundefined{definecolor}
 {\usepackage{color}}{}
\@ifundefined{definecolor}{\usepackage{color}}{}
\@ifundefined{definecolor}{\usepackage{color}}{}

\usepackage{babel}

\newcommand{\bra}[1]{\langle#1 |}
\newcommand{\ket}[1]{|#1 \rangle}
\newcommand{\braket}[2]{\left \langle #1 \mid #2 \right \rangle}
\newcommand{\ketbra}[2]{\vert #1 \rangle \! \langle #2 \vert}
\newcommand{\overlap}[2]{\left \langle #1 | #2 \right\rangle}

\newcommand{\sandwich}[3]{\left \langle #1 \mid #2 \mid #3 \right\rangle}

\usepackage{babel}

\usepackage{babel}

\usepackage{babel}

\makeatother

\usepackage{babel}
\begin{document}

\title{Non-Markovian Equilibration Controlled by Symmetry Breaking}

\author{Nicholas Chancellor, Christoph Petri, and Stephan Haas}

\address{Department of Physics and Astronomy and Center for Quantum Information
Science \& Technology, University of Southern California, Los Angeles,
California 90089-0484, USA }
\begin{abstract}
We study the effects of symmetry breaking on non-Markovian dynamics
in various system-bath arrangements. It is shown that by breaking
certain symmetries features signaling non-Markovian time evolution
disappear within a finite time $t_{g}$. We demonstrate numerically
that the scaling of $t_{g}$ with the symmetry breaking strength is
different for various types of symmetry. We provide a mathematical
explanation for these differences related to the spectrum of the total
system-bath Hamiltonian and provide arguments that the scaling properties
of $t_{g}$ should be universal. 
\end{abstract}
\maketitle

\section*{Introduction}

Non-Markovian dynamics occurs in systems connected to a bath when
information travels from the bath into the system. The name non-Markovian
arises from the fact that these systems cannot be described by Markovian
master equations \cite{Breuer(2002),Rivas(2012),Krovi(2007)}. A system
which is non-trivially joined to a \emph{finite} bath will always
be necessarily non-Markovian because any information which leaves
the system must return \cite{Boccheri(1957)} according to the Poincare
recurrence theorum. In a generic finite system, the recurrence time
is expected to scale exponentially with the size of the Hilbert space\cite{Venuti2010}.
For many systems with a moderately sized finite bath, it may therefore
be the case that significant non-Markovian events are only observable
on timescales much longer than those of other relevant dynamics.

In order to define and quantify non-Markovianity in quantum systems
different methods have been developed. For example, one can examine
a distance measure between the actual evolution and the best Markovian
model of a system bath arrangement. However, this approach is rarely
used because of the difficulty of finding an optimum Markovian approximation\cite{Wolf(2008)}.
There are other methods which involve examining the mathematical structure
of the map defining the time evolution of the system \cite{Shabani(2009),Wilkie(2009),Budnini(2006)}.
Other techniques which have been proposed are based on determining
the minimum amount of noise required to make the dynamics Markovian\cite{Wolf(2008)}
or studying the evolution of entanglement for a system prepared initially
in a maximally entangled state\cite{Rivas(2010)}.

Here we choose a method of studying the degree of non-Markovianity
similar to the one proposed in \cite{Breuer2009} and further examined
in \cite{Laine(2010),Breuer(2012)}. Its central idea is to examine
the trace distance between two different initial system states under
time evolution (for examples of other methods and techniques, both
analytic and numerical, see \cite{Krovi(2007),Wolf(2008),Breuer(2009-1),Breuer(2004),Churscinski(2010),vanWonderen(2006),Budnini(2006),Barnett(2001),Wilkie(2009),Daffler(2010),Kossakowski(2009),Shabani(2009),Rivas(2010),Breuer(2012),Wissmann2012,Piilo(2008)}).
The trace distance is defined as

\begin{equation}
D_{\psi\psi'}(t)=\mathrm{\frac{1}{2}Tr}(\sqrt{(\rho_{s}(t,\ket{\psi_{0}})-\rho_{s}(t,\ket{\psi'_{0}}))^{2}},\label{eq:traceDist}
\end{equation}

where $\ket{\psi_{0}}$ and $\ket{\psi'_{0}}$ are two different initial
states which are product states of the system and the bath with the
same bath states but orthogonal system states. The system density
matrix at a time $t$ is defined as

\begin{equation}
\rho_{s}(t,\ket{\psi_{0}})=\mathrm{Tr}_{bath}(\ketbra{\psi(t)}{\psi(t)}),\label{eq:rho_s}
\end{equation}

\[
\ket{\psi(t)}=\exp(-iH\: t)\ket{\psi_{0}},
\]

where $H$ is the overall system-bath Hamiltonian. A system is necessarily
behaving in a non-Markovian fashion if the trace distance increases
in time because of its interpretation as a measure of distinguishability.
In general, the maximum of $D_{\psi\psi'}(t)$ over all possible initial
states has to be calculated (see Ref. \cite{Breuer2009}). However,
in the case where the system has only a few degrees of freedom, a
single pair of orthogonal states can still provide valuable insights
into the overall non-Markovian behavior of the system. Furthermore,
examining non-Markovian behavior of a system by studying only two
orthogonal initial states has the advantage that in numerical studies
it is often impractical to maximize over all possible initial state
pairs.

Non-Markovian systems have been studied intensively both theoretically
and experimentally in recent years. On the experimental side the transition
between Markovian and non-Markovian dynamics has been probed in optical
setups \cite{Chiuri(2012),Liu(2011)}. These experiments have been
able to go from one regime to the other in a controllable way, thus
allowing them to control the flow of information between system and
bath. Non-Markovian effects have also been observed in solid state
systems, for example with electron transport though a quantum dot
\cite{Ubbelohde(2012)}. Theoretically, recent research in the field
of non-Markovian dynamics has been devoted to e.g. the study of memory
effects\cite{Krovi(2007),Churscinski(2010),Daffler(2010)}, including
examining the effect of bath memory on the equilibrium state of a
system after evolving for a long time\cite{Churscinski(2010)}, the
influence of a chaotic environment on non-Markovian dynamics \cite{Mata(2012)}
or the connection between non-Markovianty and entanglement \cite{Huelga(2012)}.
Interesting work has also been done in the study of multiple time
correlation functions \cite{Alonso(2005)}. We should also mention
the vast theoretical effort to understand non-Markovianity in the
framework of measurement theory\cite{Diosi(2012),Gardiner(1985),Jack(1999),Jack(2000),Diosi(2008),Wiseman(2008),Diosi(2008)-1,Mazzola(2009),Gambetta(2003)}.
Still, our knowledge on the question how the internal structure of
the bath influences the dynamics of the system is incomplete.

There have also been interesting studies on non-Markovian effects
in a structured environment, using photonic crystals \cite{Florescu(2001),John(1995)}
as well as atoms on an optical lattice coupled to matter waves \cite{de Vega(2008)}.
The work in \cite{John(1995)} is of particular relevance to this
paper, because that work examined the effects of spontaneous symmetry
breaking. This paper does not, however examine timescales of dynamical
symmetry breaking as we do.

In this paper we demonstrate that non-Markovian dynamics can be controlled
by breaking symmetries (e.g. parity) of the complete system-bath Hamiltonian.
We show that non-Markovian events in the temporal evolution of the
trace distance disappear on a certain time scale which depends on
the strength of the symmetry breaking and give spectral arguments
to show that the scaling behavior we observe is universal.

This paper is structured as follows. In Sec. \ref{sec:Setup} we explain
the two system bath Hamiltonians studied in this paper, as well as
the initial states and why they are chosen. In Sec. \ref{sec:Two-Dimensional-Tight-binding}
we look at the dynamics of a tight binding torus, which has combined
system-bath symmetry. In Sec. \ref{sec:Fully-Connected-Graph} we
look at a 2 level qubit system coupled to a completely connected tight
binding bath. In this case we break a symmetry which exists only in
the bath. This is followed by a brief summary of conclusions. More
detailed calculations are given in the supplemental material. Analysis
of the bath correlations for the system bath arrangements considered
in this paper is also provided in Sec. \ref{sub:corr} of the supplemental
material.

\section{Setup\label{sec:Setup}}

\begin{figure}
\includegraphics[scale=0.4]{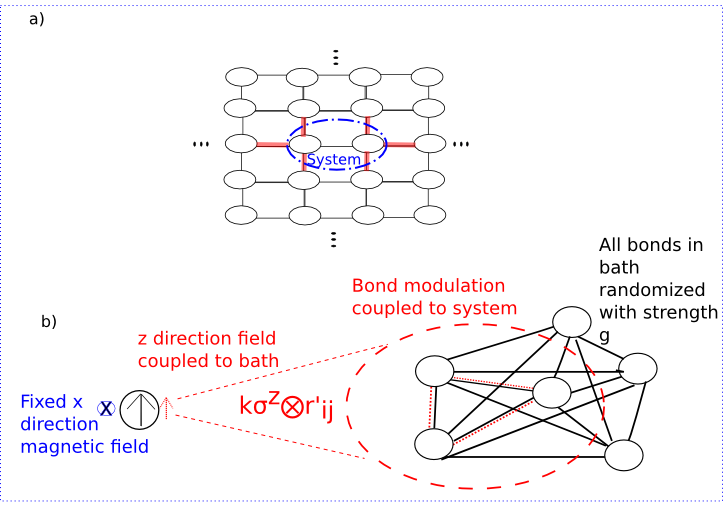}

\caption{Illustration of the two models studied in this paper a) system: 2
tight binding sites. bath: surrounding 2 dimensional tight-binding
lattice b) system: single spin-$\frac{1}{2}$ in a magnetic field,
bath: fully connected tight binding graph. Quench: system-bath coupling
turned on at t=0. \label{fig:intro}}
\end{figure}

The first setup we study is a two-dimensional tight binding lattice
whose dynamics is governed by the Hamiltonian

\begin{equation}
H_{2D}(g,N,r)=\sum_{<i,j>}^{N}c_{i}^{\dagger}c_{j}(1+g\: r_{ij}),\label{eq:2D Ham}
\end{equation}

where $c_{i}$($c_{i}^{\dagger}$) is the annihilation (creation)
operator on site \emph{i} and $N$ is the number of sites in the lattice.
The $<i,j>$ notation indicates that the sum is performed only over
adjacent sites. $r_{ij}$ is a symmetric matrix, i.e. $r_{ij}=r_{ji}$
with the additional property $|r_{ij}|<1$, whose elements are chosen
in a uniform random fashion. $g$ is a positive real number measuring
the ``strength'' of the symmetry breaking. In the case of $g=0$
we recover the Hamiltonian of a particle which can move freely between
the sites of the lattice. For finite values of $g$ the matrix elements
of the Hamiltonian are altered randomly, which corresponds to altering
the hopping rates between the sites. We initialize the system in the
ground state of the Hamiltonian where 2 sites are disconnected. In
the following these sites are referred to as the ``system'' and
the remaining sites are referred to as the ``bath''. At $t=0$ the
system and the bath are joined in a ``quench'' process. Fig. \ref{fig:intro}a)
shows the lattice and the quench. Evidently, the two eigenstates of
the two site system Hamiltonian without any connection to the bath
are a good choice of initial system states for the trace distance
analysis.

The second setup we consider is a two-level (spin $\frac{1}{2}$)
system coupled to a bath which is a fully connected graph. We have
chosen a fully connected graph as the bath Hamiltonian because it
has a very high degree of symmetry (symmetry under any permutation
of sites) and therefore is a natural choice for examining the effects
of symmetry breaking. The corresponding complete system-bath Hamiltonian
is given by

\begin{equation}
H_{conn}(g,k,N,m,r,r')=\sigma^{x}\otimes1_{bath}+\label{eq:conn Ham}
\end{equation}

\[
1_{sys.}\otimes\sum_{i,j}^{N}c_{i}^{\dagger}c_{j}(1+g\: r_{ij})+k\:\sigma^{z}\otimes\sum_{p\neq q}^{m}c_{p}^{\dagger}c_{q}r'_{pq},
\]

which is of the canonical form $H=H_{sys.}\otimes1_{bath}+1_{sys.}\otimes H_{bath}+k\: H_{sys.}^{coup}\otimes H_{bath}^{coup}$.
As before, $r'_{pq}$ and $r_{ij}$ are real symmetric matrices, ($r_{pq}'=r_{qp}'$,
$r_{ij}=r_{ji}$) with the property $|r'_{pq}|<1$ , $|r{}_{ij}|<1$.
The elements of $r'_{pq}$ and $r{}_{ij}$ are selected in a uniform
random way. k measures the strength of the coupling between the system
and the bath. For $k=0$ the system and the bath are completely decoupled.
As k is increased a coupling is generated between the direction of
the spin and the tunneling constants in the subset of sites 1 through
m of the bath. We consider the case where $0<k\leq1$. Depending on
the sign of $r'_{pq}$this coupling leads to either an increase or
decrease in the hopping strength on each bond in the subset. If the
direction of the system spin is switched, the hopping probabilities
which were more favorable will become less favorable and vice versa.
As in the first case g is a real number which measures the strength
of the symmetry breaking. Fig. \ref{fig:intro}b) shows a sketch of
this system-bath arrangement. For this Hamiltonian the spin is prepared
in the +z (-z) direction and the bath is initiated with the lower
energy quarter (25\%) of its $k=0$ eigenstates filled with non-interacting
particles. By means of this choice we can maximize the amount of information
exchange between the system and the bath (for details see Sec. \ref{sub:trace_mean}).
At $t=0$ a quench is performed when the spin-bath coupling $k$ is
turned on.

We should also briefly discuss the definition of equilibration we
use in this paper. We will consider a system to be equilibrated when
remains near a stationary state which is relatively independent of
the initial state for a long time. The trace distance measure we study
serves as an indicator of equilibration, in the sense that it shows
independence from the initial state. $D_{\psi\psi'}(t)$ however cannot
strictly be used as a measure of equilibration by itself because it
does not provide information about weather or not the long time state
of the system is a stationary state. The system is in fact equilibrating
in both cases, as we demonstrate in section \ref{sub:svn} of the
supplemental material.

\section{Two Dimensional Tight-binding Lattice with Combined System-Bath Symmetry\label{sec:Two-Dimensional-Tight-binding}}

\begin{figure}
\includegraphics[scale=0.5]{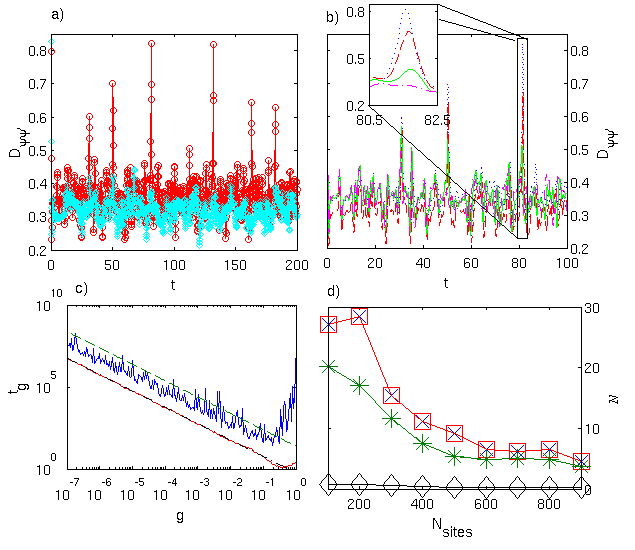}

\caption{a) Trace distance versus time for two 10x10 lattices, (red) circles
are for a torus and (cyan) diamonds are for a strip. b) Trace distance
versus time for a 10x10 torus at half filling for various strengths
of randomness g. g=0 (dotted blue line), g=0.02 (dashed red line),
g=0.05 (solid green line), g=0.1 (dot dashed purple line). c) Timescale
at which recurrences disappear, $t_{g}$ versus g for 10x10 torus
at half filling, upper solid noisy (blue) line is the mean of $t_{g}$and
the dashed (green) line is a linear fit. The lower solid (red) line
is the median of $t_{g}$and the dot-dashed (black) line the corresponding
linear fit. d) Non-Markovianity measure $\mathscr{N}(t,\Delta t;D)$
with $\Delta t=1000$ (blue) Xes are the mean of the measure for 400
realizations from $t=0$, (red) squares are the measure performed
on the mean trace distance at times from $t=0$, (green) asterisks
are the mean of the measure for times from $t=10^{8}$ and (black)
diamonds are the measure of the mean at times from $t=10^{8}$ . All
data in this plot are for small randomness, $g=10^{-6}$.\label{fig:2D_plots}}
\end{figure}

Let us now discuss how the trace distance $D_{\psi\psi'}(t)$ evolves
in time for the tight binding lattice. For a discussion concerning
the numerical details, such as how we calculate the density matrix,
we refer to the supplemental material (Sec. \ref{sub:2d density matrix}).
Fig. \ref{fig:2D_plots}a) shows $D_{\psi\psi'}(t)$ for two different
boundary conditions applied to the Hamiltonian $H_{2D}$ given by
Eq. \ref{eq:2D Ham}. For periodic boundary conditions in x (x and
y) we get a strip (torus). Since we monitor the temporal evolution
of two orthogonal initial states the trace distance is one in the
beginning of the dynamics when system and bath are joined together,
that is $D_{\psi\psi'}(t=0)=1$. Afterwards $D_{\psi\psi'}(t)$ decreases
rapidly in a short time, which is not visible in Fig. \ref{fig:2D_plots}a)
and then continues to fluctuate around a small value $D_{\psi\psi'}(t=0)\approx0.35$.
In the case of the torus, which has the greater amount of symmetry
of the two geometries, these fluctuations of the trace distance $D_{\psi\psi'}(t)$
are interrupted by pronounced peaks. By performing a thorough analysis
of the dynamics of the setup at these points in time we have found
that these peaks can be explained by partial reconstructions of states
which are similar to the initial states having a trace distance of
one. In the following we elucidate this mechanism of state reconstruction.

For the torus the outgoing wavefunction from the initial state which
is localized on the system sites propagates freely in both directions
when the connections between the system and the bath are turned on.
Eventually, it returns coherently to its initial position, reconstructing
a wavefunction which is mostly isolated within the system. Similarly,
the wavefunctions of the other states which initially have zero amplitude
in the system reconstruct states with an approximately similar shape.
On the contrary, for the strip geometry the wavefunctions are scattered
off the open boundary such that they do not reconstruct coherently
when they return. Formally, the torus has a $\mathbb{Z}_{N}\otimes\mathbb{Z}_{N}$
symmetry group whereas the strip only has the cyclic symmetry group
$\mathbb{Z}_{N}$\cite{Kosmann-Schwarzbach (2010}. The larger symmetry
group of the torus allows the wavefunction to preserve more of its
shape while propagating. Therefore the torus allows for reconstructions
which do not occur for the strip. Note that these reconstructions
occur at a much shorter timescale than the recurrences which must
occur because any finite closed quantum system must eventually return
to a state arbitrarily close to its initial state\cite{Boccheri(1957)}.
To avoid confusion we will refer to the features seen in Fig.\ref{fig:2D_plots}
a) as reconstructions and the features which must necessarily appear
at long times as recurrences. Recurrences occur in the strip as well
as any other geometry we could consider, but they occur at a much
longer timescale than the partial reconstructions.

The coherence of the returning waves which partially reconstruct the
initial state can be destroyed over time if randomness is introduced
to the Hamiltonian. Fig. \ref{fig:2D_plots}b) shows the temporal
evolution of the trace distance $D_{\psi\psi'}(t)$ in the case of
the torus-geometry for different values of g, the strength of the
random term. As the magnification in Fig. \ref{fig:2D_plots}b) reveals
the height of reconstruction peaks decreases with increasing $g$.
In order to analyze this behavior we have studied how the time it
takes for these peaks to disappear $t_{g}$ depends on the strength
of the randomness $g$. Fig. \ref{fig:2D_plots}c) shows $t_{g}$
as a function of $g$ in a double-logarithmic plot. As we can see
from Fig. \ref{fig:2D_plots}c) $t_{g}$ depends on $g$ in an inverse
linear fashion, i.e.

\begin{equation}
t_{g}\propto\frac{1}{g}.\label{eq:tg_taurus}
\end{equation}

In the following we derive an intuitive understanding for this behavior.
Evidently, the Hamiltonian of the complete system-bath arrangement
has a discrete translational symmetry which leads to level-degeneracy
in the energy spectrum. The addition of randomness breaks this symmetry
and therefore leads to level splitting. For small randomness the gaps,
which are created by the level splitting, are linearly proportional
to $g$. The timescale associated with these gaps is simply their
inverse which yields the dependency of $t_{g}$ on $g$ given by Eq.
(\ref{eq:tg_taurus}). The disappearance of the reconstruction by
introducing randomness to the system-bath Hamiltonian peaks can be
thought of as an equilibration process. We find that away from the
reconstruction peaks the system density matrix is close to the fully
equilibrated state for both initial conditions. The reconstructions
drive the system away from this state, and therefore the system becomes
more equilibrated as they are destroyed by symmetry breaking. The
timescale $t_{g}$ with which these peaks disappear can therefore
be thought of as a timescale for the equilibration of the system.
(We also note that in Fig. \ref{fig:2D_plots}c) the mean of $t_{g}$
is quite noisy and is much greater than the median. This indicates
the presence of rare events where $t_{g}$ is very large.) In this
case $t_{g}$ is defined as the timescale on which the reconstruction
peaks no longer occur. Different $t_{g}$ can be observed for different
choices of $r_{ij}$ from Eq.\ref{eq:2D Ham}. If we call $t_{g}$
for the kth random choice of $r_{ij}$ $t_{g}^{(k)}$, than the mean
will be $t_{g}^{mean}=\sum_{k=1}^{m}\frac{t_{g}^{(k)}}{m}$. If the
times are ordered such that $t_{g}^{(k)}\leq t_{g}^{(k+1)}$ than
for m total samples, the median is $t_{g}^{median}=t_{g}^{(\frac{m}{2})}$.

Finally, we examine the effect of the reconstructions on the non-Markovianity
of the system. To this end we calculate the total increase in trace
distance over a given time period defined as

\begin{equation}
\mathscr{N}(t,\Delta t,D)=\intop_{t}^{t+\Delta t}d\tau[\frac{\partial D_{\psi\psi'}(\tau)}{\partial\tau}\Theta(\frac{\partial D_{\psi\psi'}(\tau)}{\partial\tau})],\label{eq:nm_measure}
\end{equation}

where $\Theta$ is the Heavyside function and $D_{\psi\psi'}(\tau)$
is defined in Eq. \ref{eq:traceDist}. $\mathscr{N}(t,\Delta t,D)$
is a measure of the information flowing from the bath into the system
which allows us to distinguish between $\psi$ and $\psi'$. It can
therefore be thought of as the information allowing us to distinguish
between $\psi$ and $\psi'$ which would necessarily be lost by a
Markovian approximation of the bath. Any time distinguishably between
two arbitrary states is increasing the behavior of the system can
be said to be non-Markovian\cite{Breuer2009}. Fig. \ref{fig:2D_plots}d)
shows $\mathscr{N}(t,\Delta t,D)$ as a function of the number of
system sites $N_{sites}$ at different points in time t with a fixed
$\Delta t$. At early times the measure is the same for all realizations
because the symmetry is not broken yet so the mean of the measure
is the same as the measure of the mean. We see that with increasing
time $\mathscr{N}(t,\Delta t,D)$ decreases systematically for all
system sizes. Consequently, the evolution is less non-Markovian after
the reconstruction peaks have disappeared. Furthermore, at times after
the symmetry breaking has occurred, the average over different realizations
of the system and bath does not show any increase. This indicates
that any remaining non-Markovian features at these late times depend
strongly on the details of the random symmetry breaking term. The
fluctuations observed at these times are therefore not intrinsic dynamical
fluctuations of the system, but a result of finite size and the presence
of random symmetry breaking terms. Since the coupling between the
system and bath is quite strong, there is no reason to expect Markovianity
even asymptotically and in the large bath limit.

Moreover, Fig. \ref{fig:2D_plots}d) reveals that as the system size
$N_{sites}$ becomes larger the early and late non-Markovianity measures
approach each other. An explanation for this is that as the system
is made larger there is more time between the reconstructions. Thus
the effect of the reconstructions is hidden by small fluctuations
which transport information in and out of the system at all times.

\section{Fully Connected Graph Coupled to 2 Level System\label{sec:Fully-Connected-Graph}}

\begin{figure}
\includegraphics[scale=0.5]{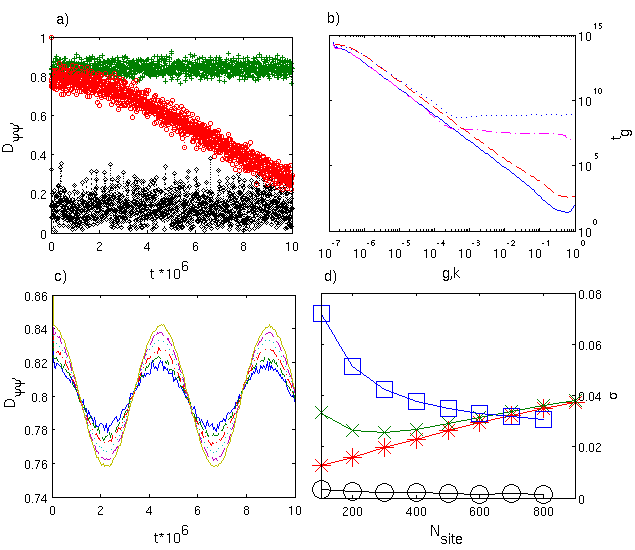}

\caption{a) Trace distance between two initially orthogonal states versus time
for a bath consisting of 25 particles on 100 sites (N=100) with k=1
and m=20. Pluses (green) correspond to $g=0$, circles (red) to $g=10^{-3}$,
and diamonds (black) to $g=10^{-1}.$ b) $t_{g}$ versus the strength
of the randomness $g$ respectively the coupling parameter between
system and bath $k$. \textcolor{black}{The solid (blue) line is $t_{g}$
versus g for the same parameters as (a), the dot-dashed (magenta)
line is the same but with $k=10^{-3}$. The dashed (red) line is $t_{g}$
versus k for g=1 and all other parameters the same as (a), the dotted
(green) line is the same but with $g=10^{-3}$. }c) $D_{\psi\psi'}(t)$
for various system sizes with $k=1$, $m=\frac{N}{5}$, and quarter
filling. Solid (blue) line is for $N=100$, dashed (green) $N=200$,
dot-dashed (red) $N=300$, dotted (cyan) $N=400$, dashed (purple)
$N=500$ and solid (gold) $N=600$. All plots are averaged over 400
realizations of the system-bath arrangement. d) Standard deviation
of mean trace distance averaged over 400 samples for $g=10^{-4}$.
Standard deviation of mean (red asterisks) and mean of standard deviation
(green Xes) for times from $t=10^{5}$ to $t=10^{7}$, standard deviation
of mean (black circles) and mean of standard deviation (blue squares),
for times from $t=10^{9}$ to $t=10^{11}$.\label{fig:conn_plots}}
\end{figure}

Let us now turn to the second setup which is a qubit coupled with
a fully connected tight binding graph. Fig. \ref{fig:conn_plots}a)
shows the temporal evolution of the trace distance $D_{\psi\psi'}(t)$
for different values of the randomness $g$. For $g=0$ we observe
fluctuations of $D_{\psi\psi'}(t)$ around $0.8$ after a rapid decrease
from the initial value $D_{\psi\psi'}(t=0)=1$. When the randomness
is large, $g=10^{-1}$, the fluctuations occur at a much smaller value,
that is $D_{\psi\psi'}(t)\approx0.2$. For moderate randomness, $g=10^{-3}$,
the trace distance slowly decreases in time until it continues to
fluctuate around an intermediate value between $0.2$ and $0.8$.
Accordingly, it can be said that the system equilibrates to states
with a much larger trace distance for $g=0$ compared to the case
when $g$ is finite (see sec. \ref{sub:svn} of the supplementary
materials for more information on why this can be considered equilibration)
. In other words one can say that the unbroken permutation symmetry
for $g=0$ prevents full equilibration for which we give a physical
explanation in the next paragraph.

For $g=0$ the high degree of symmetry leads to many completely localized
eigenstates in the part of the bath which is not connected to the
system. Consequently, these parts of the wavefunction cannot contribute
to the equilibration process. Hence for $g=0$ the system will never
fully equilibrate. When a small randomness ($g\neq0$) is added, these
eigenstates acquire a finite amplitude to be in the part of the bath
connected to the system and thus the system will eventually equilibrate.
(see Sec. \ref{sub:full_spect})

Contrary to the first setup we observe in Fig. \ref{fig:conn_plots}a)
no peaks or systematic increases of the trace distance $D_{\psi\psi'}(t)$
which would correspond to a flow of information from the bath into
the system. Consequently, one could draw the conclusion that the second
setup shows no systematic non-Markovian dynamics which do not depend
on the details of $r'_{pq}$ and $r_{ij}$. However, this is not true
and we will return to this problem at a later point in the manuscript.

Let us now ask how the strength of the random term $g$ affects the
equilibration time $t_{g}$ which we define as the time when the trace
distance drops below $0.4$. In Fig. \ref{fig:conn_plots}b) $t_{g}$
versus $g$ is shown in a double logarithmic plot. As we can see the
equilibration time scales as

\begin{equation}
t_{g}\propto\frac{1}{g^{2}}.\label{eq:tg_conn}
\end{equation}

as long as $g$ is sufficiently small compared to the coupling parameter
$k$. Furthermore, we observe that if $g$ is large compared to $k$
the equilibration time $t_{g}$ does not depend on $g$.

The physical reason for the different dependencies of the equilibration
time $t_{g}$ on the strength of the randomness $g$ for the two studied
setups (compare Eqs. \ref{eq:tg_taurus} and \ref{eq:tg_conn}) can
be traced back to the spectrum of the system bath Hamiltonian. Both
Hamiltonians show a level splitting which is proportional to $g$.
Nevertheless, this level splitting leaves a gap degeneracy in the
case of the fully connected bath by gap degeneracy we mean that the
gaps between certain energy eigenvalues of the overall Hamiltonian
are the same, in this case the gap degeneracy is the result of pairs
of energy levels having the same slope when the level degeneracy is
split. We will later show that this gap degeneracy is effectively
the same as a level degeneracy once we trace out the bath degrees
of freedom. A demonstration that this happens whenever there is a
symmetry in the bath which is not broken by coupling to the system
can be found in Sec. \ref{sec:spect_args}.

As the next step we keep the randomness fixed and study how the equilibration
time $t_{g}$ depends on the strength of the coupling between the
system and the bath $k$. Fig. \ref{fig:conn_plots} b) shows that
$t_{g}$ scales exactly as before in an inverse quadratic way, i.e.

\begin{equation}
t_{g}\propto\frac{1}{k^{2}},\label{eq:tg_conn1}
\end{equation}

which we are going to explain in the following. In this case the tensor
product nature of the k=0 Hamiltonian produces a gap degeneracy. The
tensor product structure of the coupling term in Eq. \ref{eq:conn Ham}
means that this gap degeneracy cannot be broken by direct level splitting,
but instead must be broken by level repulsion, for more details see
Sec. \ref{sub:t_g-k}.

Now we return to question whether the dynamics of the second setup
possesses hallmarks of systematic non-Markovianty or not. To do this
we must compare the mean of the standard deviation of the trace distance,
which shows all fluctuations, both intrinsic dynamical fluctuations
and those related to the random symmetry breaking terms which is defined
as 

\begin{equation}
M_{\sigma}=\sum_{k=1}^{s}\frac{1}{s}\sqrt{\sum_{i=1}^{p}\frac{1}{p}(D_{\psi\psi'}^{(k)}(t_{i})-\sum_{i=1}^{p}\frac{1}{p}D_{\psi\psi'}^{(k)}(t_{i}))^{2}},
\end{equation}

to the standard deviation of the mean

\begin{equation}
\sigma_{M}=\sqrt{\sum_{i=1}^{p}\frac{1}{p}(\sum_{k=1}^{s}\frac{1}{s}D_{\psi\psi'}^{(k)}(t_{i})-\sum_{k=1}^{s}\frac{1}{s}\sum_{i=1}^{p}\frac{1}{p}D_{\psi\psi'}^{(k)}(t_{i}))^{2}},
\end{equation}

which only shows the intrinsic dynamical fluctuations.

We observe that Fig. \ref{fig:conn_plots}a) shows no systematic increase
of the trace distance $D_{\psi\psi'}(t)$. Fig. \ref{fig:conn_plots}c)
shows an average of $D_{\psi\psi'}(t)$ over 400 different realizations
of $r'_{pq}$ versus the bath size whereas the fraction $\frac{m}{N}$
is fixed and we choose $k=1$, $g=0$. Evidently, we observe systematic
non-Markovian oscillations in Fig. \ref{fig:conn_plots}c) which are
masked by random fluctuations in Fig. \ref{fig:conn_plots}a). Indeed
these oscillations are systematic in the sense that they are not destroyed
by averaging over different realizations. Increasing the bath size
appears to actually increase the strength of the systematic oscillations.
These oscillations are caused by internal bath dynamics where the
bath sites which are not connected to the system effectively act as
a single site, detailed calculations of how this happens can be found
in Sec. \ref{sub:full_spect}.

After the system has fully equilibrated the non-Markovian oscillations
of the trace distance disappear as we can see from Fig. \ref{fig:conn_plots}d)
in which we examine the standard deviation of the trace distance at
early (after the initial transient but before the symmetry is broken)
and late (after the symmetry is broken) times. This can be seen because
at late times the standard deviation of the mean over different choices
of $r_{ij}$ and $r'_{pq}$ drops to nearly zero. The drop to nearly
zero means that the systematic oscillations have disappeared because
the mean value over different realizations hardly fluctuates at all.
At these late times we can see that there are fairly strong fluctuations
in $D_{\psi\psi'}(t)$, these fluctuations disappear however when
we average over different realizations. The analogue of Fig. \ref{fig:conn_plots}c)
at late rather than early times would look like a flat line and not
contain the sinusoidal oscillations we see at early times.

By examining the standard deviation of the mean trace distance compared
to the mean of the standard deviation we can determine how sensitive
the non-Markovian behavior is to the specific values of the randomly
selected term $r'_{pq}$ (and for late times also $r_{ij}$). As Fig.
\ref{fig:conn_plots}d) shows the standard deviation of the mean of
trace distance at early times increases and is approached by the mean
of the standard deviation for large system size. The closeness of
these two numbers indicate that for large bath sizes the trace distance
at early times is dominated by behaviors which are the same for most
choices of $r'_{pq}$ and do not depend much on the details of the
random modulation of bond strengths. In contrast the standard deviation
of the mean trace distance for late times is much less than the mean
of the standard deviation for all system sizes, indicating that asymptotically
the trace distance is dominated by random fluctuations strongly dependent
on the details of $r_{ij}$ and $r'_{pq}$.

In Fig. \ref{fig:conn_plots}d) we examine the mean of the standard
deviation of the trace distance over different random selections of
$r_{ij}$ and $r'_{pq}$, we also examine the standard deviation of
the mean value of trace distance. The mean of the standard deviation
tells us about the average variability of the trace distance before
and after the symmetry is broken. This mean decreases with system
size for late times and eventually becomes less than the value for
early times for a bath with 800 total sites. This decrease with size
is caused by a decrease in non-Markovian fluctuations associated with
finite size effects as the bath is made bigger. In contrast the mean
of the standard deviation for early times increases with system size
as the trace distance oscillations shown in Fig. \ref{fig:conn_plots}c)
get stronger.

\section{Spectral Arguments for the Universality of $t_{g}\propto\frac{1}{g^{2}}$
For bath symmetry breaking\label{sec:spect_args}}

\begin{figure}
\includegraphics[scale=0.4]{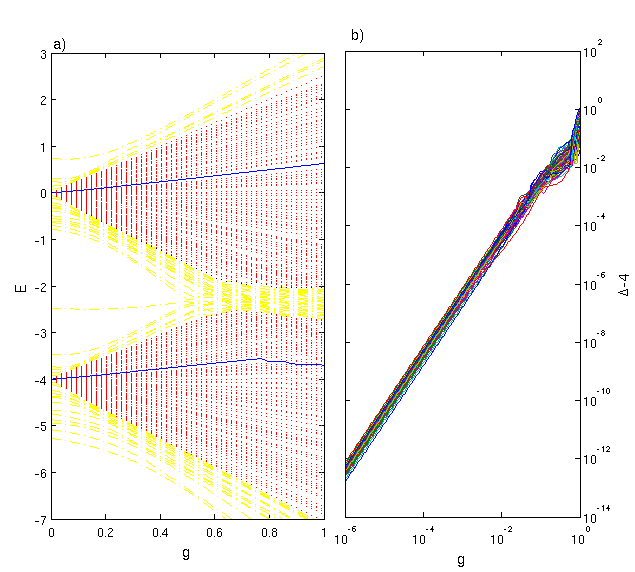}

\caption{(color online) a) Energy spectrum of the Hamiltonian for various strengths
of g, dot-dashed (yellow) lines are eigenstates which are not initially
part of a degenerate manifold of states for g=0, dotted (red) lines
are states which are part of a degenerate manifold for g=0. A pair
of eigenstates which are the same in the bath for g=0 are shown as
sold (blue) lines. The slope of the energy of these two states is
the same for small g but is changed by avoided crossings for larger
g. b) Double logarithmic plot of the difference from the initial gap
between pairs of states within the degenerate manifold which are the
same in the bath for g=0 versus g. \label{fig:gap_deg}}
\end{figure}

An explanation for why $t_{g}$ scales like $\frac{1}{g^{2}}$ rather
than $\frac{1}{g}$ can be obtained by examining the energy spectrum
of the Hamiltonian at different values of g. As we can see from Fig.
\ref{fig:gap_deg}a), the spectrum for g=0 is highly degenerate. As
g is increased, the degeneracy is broken, however there will be pairs
of states within the two manifolds which change energy with the same
slope, as we can see from Fig. \ref{fig:gap_deg}a). The gap between
the energies of the states in these pairs will therefore remain the
same until the slope of the energy is changed by an avoided crossing.
The relevant process for equilibrating the system is not the splitting
of the \emph{level }degeneracy but rather the splitting of a \emph{gap}
degeneracy caused by avoided crossings. Because the slope is the same
for each pair, the gap must change in a manner proportional to $g^{2}$
rather than g. We can verify that the gap increases proportionally
to $g^{2}$ by looking at the double logarithmic plot of gap versus
g shown in Fig. \ref{fig:gap_deg}b).

The two degenerate manifolds in \ref{fig:gap_deg}a) correspond to
states where the particle in the bath is isolated in the sites which
are not connected to the system. These two manifolds correspond to
the two eigenstates of the system Hamiltonian. The total Hamiltonian
can therefore be written in the form

\begin{equation}
H=(H'{}_{sys}\otimes1_{sym})\oplus H_{nosym}+g\:1_{sys}\otimes H_{break}.\label{eq:Hsym}
\end{equation}

This is the generic form of any system bath arrangement with a bath
symmetry. In general we can always write $H_{break}=\lambda_{split}(H_{split}\oplus H_{2})+\lambda_{mix}H_{mix}$
where $H_{split}$ spans the same subspace as $1_{sym}$ in Eq.\ref{eq:Hsym}.
The Hamiltonian is now of the form

\begin{eqnarray}
H & = & (H'{}_{sys}\otimes1_{sym}+g\:\lambda_{split}\:1_{sys}\otimes H_{split})\label{eq:Hsplit}\\
 &  & \oplus(H_{nosym}+g\:\lambda_{split}\: H_{2})+g\:\lambda_{mix}\:1_{sys}\otimes H_{mix}.\nonumber 
\end{eqnarray}

We can now construct a basis in the system such that $[\ket{n}\bra{n},H'_{sys}]=0\;\forall\ket{n}$
and a basis in the bath such that $\bra{i}1_{sym}\ket{i}=\begin{cases}
1 & \ket{i}\in deg.\\
0 & otherwise
\end{cases}$ where deg. refers to the degenerate subspace created by the bath
symmetry. For $\lambda_{mix}=0$ the density matrix of a system with
an arbitrary initial state $\ket{\psi_{0}}$ can then be written as

\begin{eqnarray}
\rho_{ijmn}(t) & = & \exp(-i\Delta_{mn}t)\label{eq:rho_split}\\
 & \cdot & (\sum_{l,k\in sym}A_{ijlkmn}\exp(-i\Delta_{ij}t))\nonumber \\
 & + & \sum_{l\: or\: k\notin sym}A{}_{ijlkmn}\exp(-i\Delta{}_{lk}t),\\
A_{ijlkmn} & = & (\bra{m}\otimes\bra{i})\ket{l}\overlap{l}{\psi_{0}}\overlap{\psi_{0}}{k}\bra{k}(\ket{j}\otimes\ket{n}),\nonumber \\
\Delta_{lk} & = & E_{l}-E_{k}\nonumber 
\end{eqnarray}

Where $\ket{l}$ and $\ket{k}$ are eigenstates of the overall Hamiltonian
and $\Delta_{mn}=\sandwich{m}{H'_{sys}}{m}-\sandwich{n}{H'_{sys}}{n}$
is the gap between states within the different degenerate manifolds
which is determined uniquely by n and m. We can now trace out the
bath, yielding the system density matrix,

\begin{eqnarray}
\rho_{nm}(t) & = & \sum_{q}(\exp(-i\Delta_{nm})(\sum_{l,k\in sym}A_{qqlkmn})\label{eq:rhoR_split}\\
 & + & \sum_{l\: or\: k\notin sym}A{}_{qqlkmn}\exp(-\imath\Delta{}_{lk}t))\nonumber \\
 & = & \sum_{q}\sandwich{q}{\rho(t)}{q}.\nonumber 
\end{eqnarray}

Notice that all of the terms with $\Delta_{ij}\neq0$ do not contribute
to the final reduced density matrix, and therefore the sector of the
bath containing $H_{split}$ cannot dephase the system. If $\lambda_{mix}\neq0$
than level repulsion can occur with energy levels outside of the subspace
as shown in Fig. \ref{fig:gap_deg}, these level repulsions break
the gap degeneracy and therefore destroy the tensor product structure
in Eq. \ref{eq:Hsplit}. As we can see from Fig. \ref{fig:gap_deg}b)
the gaps are widened proportionally to $g^{2}$ rather than g, therefore
the timescale for the broken symmetry to equilibrate the system is
proportional to $\frac{1}{g^{2}}$. The arguments given here will
work for any case where a bath symmetry is broken, therefore for small
g, $t_{g}\propto\frac{1}{g^{2}}$ for a generic bath symmetry being
broken%
\footnote{Technically there are ways to mathematically construct Hamiltonians
for which $t_{g}\propto\frac{1}{g^{n}}$ where $n>2$, but these Hamiltonians
are a zero measure set in the sense that a generic perturbation to
the bath will restore $t_{g}\propto\frac{1}{g^{2}}$.%
}.

To make this argument more concrete, let us consider the case where
we do not constrain the form of $H_{break}$, we can always write
$H_{break}=(H_{spit}\oplus H_{2})+H_{mix}$, we note that only $H_{mix}$can
break the gap degeneracy. We can further note that in this way of
writing $H_{break}$ we can always choose $H_{mix}$ in such a way
that it contains only off diagonal terms which mix between the sectors
of the bath Hamiltonian with and without the degeneracy.

The only effect $H_{mix}$ can have on the energy levels within the
degenerate subspace is though level repulsion with levels outside
of this subspace (shown as dashed (yellow online) lines in Fig. \ref{fig:gap_deg}
a). Let us consider the effect where 2 energy levels are initially
separated by a gap $\Delta_{0}$ and an off diagonal term $g_{0}$
is added to split the degeneracy. For simplicity let us choose the
zero of energy to lie half way between the 2 levels such that $E_{0}^{1}=\frac{\Delta_{0}}{2}=-E_{0}^{2}$.
In this case we can easily find the energy as a function of $g_{0}$
\begin{equation}
E_{g}^{1}=\sqrt{\frac{\Delta_{0}^{2}}{4}+g_{0}^{2}}=\frac{\Delta_{0}}{2}+\frac{2g_{0}^{2}}{\Delta_{0}}+O(g_{0}^{4})\label{eq:level_splitting}
\end{equation}

Because $g_{0}\propto g$ this implies that for small values of g
the addition of a term $g\: H_{mix}$will cause the energies to deviate
from their initial values in a manner proportional to $g^{2}$. The
time scale associated with this modification of the spectrum will
be inversely proportional to the splitting of the gap degeneracy,
therefore $t_{g}\propto\frac{1}{g^{2}}$.

The case where more than 2 (but still finitely many) energy levels
are simultaneously involved in level repulsion is slightly more complicated.
In this case we can argue that because the energy is an analytic function
of g for all finite (and zero) g, it can be represented as a perturbation
series expanded around g=0,

\begin{equation}
E_{g}^{1}=a_{0}+a_{1}g+a_{2}g^{2}+\ldots+a_{\alpha}g^{\alpha}+\ldots.\label{eq:Taylor series}
\end{equation}
 The direct level splitting will always be the same for all sets of
eigenvalues with the same bath state, and the energy levels will be
initially degenerate. We have already show that $H_{split}$ cannot
play a role in the equilibration. We can now further deduce that the
first order contribution to the perturbation series from $H_{mix}$
vanishes because $\sandwich{\phi_{split}}{H_{mix}}{\phi_{split}}=0$
where $\ket{\phi_{split}}$ is an eigenvector of $H_{split}$. For
this reason the constant and linear terms will not play a role in
the equilibration. It is mathematically possible that the level repulsions
are precisely balanced so that the term $a_{2}$ in Eq. \ref{eq:Taylor series}
is equal to zero, in fact with enough energy levels it is mathematically
possible to make all terms from $a_{2}$ up to $a_{\alpha-1}$, where
$\alpha$ is a finite integer%
\footnote{Consider for example the case where $E_{0}$ is involved with level
repulsion with a set of energy levels $E_{i}$ which have off diagonal
terms $k_{i}\, g$ between themselves and $E_{0}$ but do not have
any between each other. In this case the $a_{n}$ from Eq. \ref{eq:Taylor series}
can be calculated exactly and are $a_{0}=E_{0},\quad a_{n}=\frac{(-1)^{\frac{n}{2}}}{2}(1+(-1)^{n})\frac{\Gamma(\frac{3}{2})}{\Gamma(\frac{n}{2}+1)\Gamma(\frac{3}{2}-\frac{n}{2})}_{i}\sum_{i}\frac{\textrm{sign}(\Delta_{i})|k_{i}|^{n}}{|\Delta_{i}|^{n-1}}$
where $\Delta_{i}=E_{0}-E_{i}$, in this simple case for m carefully
chosen $E_{i}$ and $k_{i}$ we could potentially have an $\alpha$
as large as 2m.%
}, be equal for all $E_{g}^{i}$ in the initially degenerate subspace.
In this case we would have $t_{g}\propto\frac{1}{g^{\alpha}}$, such
a case however is not typical in the sense that the repulsion strengths
between levels would have to be precisely chosen for these terms in
the perturbation series to be equal, and a small generic modification
of the symmetry breaking term would restore the more universal behavior
of $t_{g}\propto\frac{1}{g^{2}}$. Stated differently, for a given
system bath arrangement, the set of symmetry breaking terms where
$t_{g}$ does not scale like $\frac{1}{g^{2}}$ is a subset of measure
zero of the set of all possible symmetry breaking terms which have
the required bath symmetry.

\section*{Conclusions}

In this work it has been shown that non-Markovian dynamics can be
evoked by symmetry breaking. We have examined two examples of atypical
equilibration which involve highly non-Markovian behavior for a period
of time. In both of these examples, internal bath degrees of freedom
play an important role. We have shown that these non-Markovian features
are related to symmetries in the Hamiltonian, and that breaking these
symmetries determines the timescales of their disappearance. Furthermore
we have demonstrated that the effect of the symmetry breaking process
is fundamentally different for symmetries of the overall system-bath
Hamiltonian, compared to the breaking of symmetries which exist only
in the bath, but are not broken by coupling to the system. There is
a difference in the scaling of the equilibration timescale with the
strength of the symmetry breaking. In the case of combined system-bath
symmetry breaking this time scales inverse linearly with symmetry
breaking strength, whereas when a symmetry only of the bath is broken,
it scales inverse quadratically. This difference arises from the fact
that for the breaking of a combined system-bath symmetry the relevant
equilibration timescale is related to the breaking of \emph{level
} degeneracy, whereas in the case where a bath symmetry is broken
the equilibration timescale relates to the breaking of \emph{gap}
degeneracies, in which the spacing between certain energy levels is
the same. Because of the underlying spectral cause, these types of
scaling are expected to be universal.

\section*{Acknowledgements}

The authors would like to thank Tameem Albash and Paolo Zanardi for
many helpful discussions. The numerical computations were carried
out on the University of Southern California high performance supercomputer
cluster. This research is partially supported by the ARO MURI grant
W911NF-11-1-0268.

\section{Supplemental Material\label{sec:Appendix}}

\subsection{Bath correlations for both systems\label{sub:corr}}

We can now examine the bath correlations, $|\braket{B(0)}{B(t)}|$
for the two system bath arrangements considered in this paper. In
the case of the torus, it is not immediately obvious how bath correlations
should be defined, because there is not and underlying tensor product
structure. In this case we will define B to be the part of the Hamiltonian
which is turned on at t=0 and joins the system and bath. 

\begin{figure}
\includegraphics[scale=0.4]{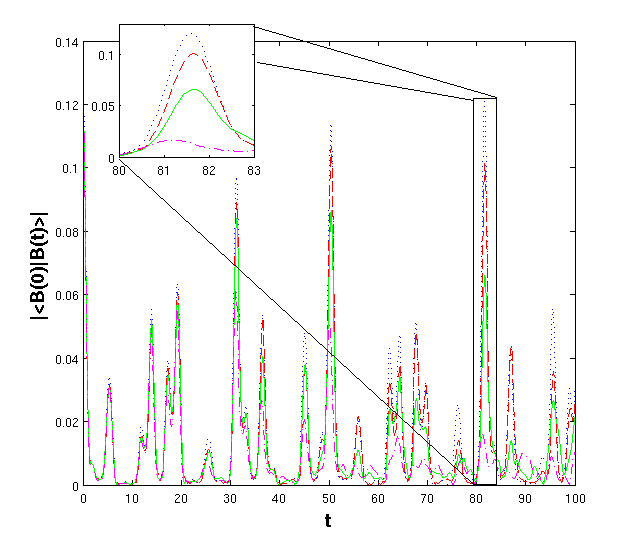}

\caption{\label{fig:tor-corr}(Color online) $|\braket{B(0)}{B(t)}|$ versus
time for 10x10 torus with the system initialized in a singlet state,
and the bath initialized at half filling. Dotted (blue) line is for
g=0, dashed (red) line is for g=0.01, solid (green) line is g=0.05
and dot-dashed (magenta) line is for g=0.1.}
\end{figure}

From Fig. \ref{fig:tor-corr} we can see that the bath correlations
of the torus system bath arrangements show the same features that
we observed in the trace distance. We notice peaks in this correlation
which disappear on a timescale governed by g. We can also notice that
the time for the bath correlations to initially decay is roughly 1
in units of coupling energy, and is independent of g.

\begin{figure}
\includegraphics[scale=0.4]{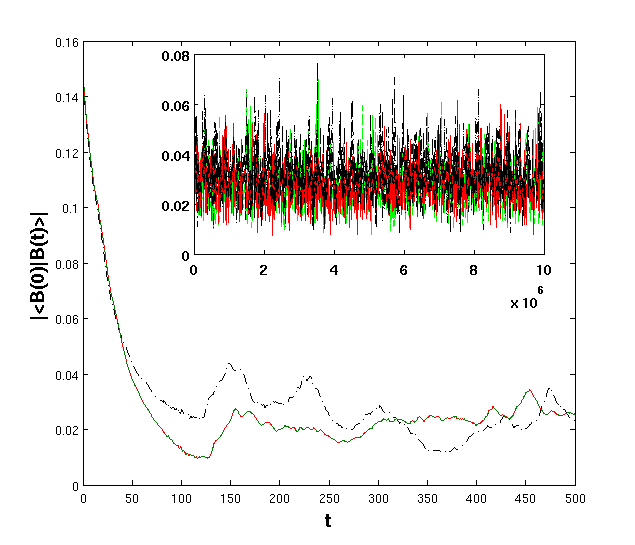}

\caption{\label{fig:conn-corr}(color online)$|\braket{B(0)}{B(t)}|$ versus
time for the fully connected arrangement for a bath consisting of
25 particles on 100 sites (N=100) with m=20. Dashed (green) lines
correspond to $g=0$, solid (red) to $g=10^{-3}$, and dot-dashed
(black) to $g=10^{-1}.$ Note that the lines for $g=0$ and $g=10^{-3}$
completely overlap throughout the main plot. Inset is the same quantities
plotted over a longer timescale.}
\end{figure}

We can now also examine the bath correlations for the fully connected
tight binding bath, in this case we define $B=\sum_{p\neq q}^{m}c_{p}^{\dagger}c_{q}r'_{pq}$.
From the inset of Fig. \ref{fig:conn-corr} we notice that the effect
of the symmetry breaking is not visible from the bath correlations.
We can also see from the main figure that the correlation decay time
is roughly the same for all values of g and is on the order of $10^{2}$
in units of the coupling energy. The first revival of the bath correlations
occurs on a similar timescale, and also appears not to depend on g.
The relative independence of the details of the bath correlations
on g indicates that the symmetry breaking effects observed in this
system-bath arrangement are not well captured by the usual method
of examining bath correlations.

\subsection{Von Neumann entropy of systems\label{sub:svn}}

As mentioned in the main text, showing that the trace distance becomes
small does not rigorously demonstrate that the system is equilibrating.
To show equilibration we need to demonstrate also that the system
is near a stationary state. Trace distance does not allow us to access
this information, but fortunately we can demonstrate this by examining
the von Neumann entropy of one of the evolving states in each of the
systems. 

We choose to use base 2 von Neumann entropy, which has the formula

\begin{equation}
S_{VN}=\textrm{Tr}(\rho\log_{2}(\rho)).\label{eq:svn}
\end{equation}

The entropy in Eq. \ref{eq:svn} is uniquely maximized by the totally
dephased state $\rho_{dep}=\frac{1_{n}}{n}$ where n is the size of
the system Hilbert space, in this case $S_{VN}^{max}=\log_{2}(n)$.
We can clearly see that $\rho_{dep}$ will commute with any unitary
time evolution operator and is therefore a stationary state. Therefore
if $S_{VN}$ is close to $S_{VN}^{max}$than the system is close to
a stationary state, thus fulfilling that criteria for equilibration.
It is important to note that the converse is not true, stationary
states do not necessarily have high $S_{VN}$, for example a pure
density matrix made from an eigenstate of the Hamiltonian is stationary,
but has $S_{VN}=0$.

\begin{figure}
\includegraphics[scale=0.4]{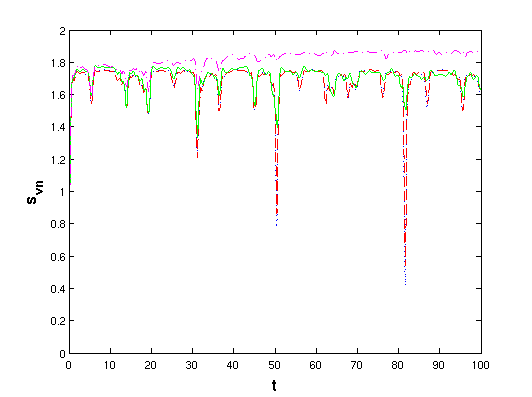}

\caption{\label{fig:svn_torus} (Color online) Von Neumann entropy versus time
for 10x10 torus with the system initialized in a singlet state, and
the bath initialized at half filling. Dotted (blue) line is for g=0,
dashed (red) line is for g=0.01, solid (green) line is g=0.05 and
dot-dashed (magenta) line is for g=0.1. The maximum possible entropy
for this system is 2.}
\end{figure}

Allow us to first examine the von Neumann entropy for the torus arrangement,
which can be seen in Fig. \ref{fig:svn_torus}. We first note that
because the system consists of 2 tight binding sites, the system density
matrix is 4x4, meaning that the maximum $S_{VN}$ is 2. We see that
except for during a reconstruction, this quantity is close to its
maximum, indicating that the system is in fact near a stationary state.
We can further notice from the plot for g=0.1 that the von Neumann
entropy will increase slightly after the reconstructions have disappeared,
this provided further evidence of an equilibration timescale related
to the symmetry breaking.

\begin{figure}
\includegraphics[scale=0.4]{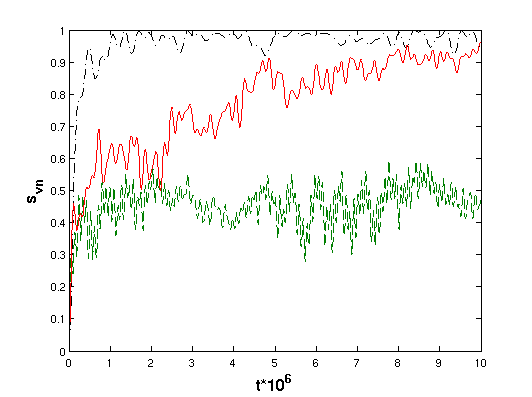}

\caption{\label{fig:svn_conn} Von Neumann entropy for the fully connected
arrangement for a bath consisting of 25 particles on 100 sites (N=100)
with k=1 and m=20. Dashed (green) lines correspond to $g=0$, solid
(red) to $g=10^{-3}$, and dot-dashed (black) to $g=10^{-1}.$ The
maximum possible entropy for this system is 1.}
\end{figure}

We now wish to perform the same analysis in the fully connected case.
By examining Fig. \ref{fig:svn_conn} we can clearly see that at times
greater than $t_{g}$ the von Neumann entropy approaches the maximum
possible value of 1 for a single spin-$\frac{1}{2}$. We therefore
can also conclude that the second system approaches a stationary state,
and therefore equilibrates.

\subsection{Details of density matrix construction for 2 dimensional graph \label{sub:2d density matrix}}

In this case we are considering a bosonic system and producing a reduced
density matrix which combines the matrix elements for measuring a
single particle on a site with those of measuring any finite (non-zero)
number of particles on that site. For a two site system the density
matrix for an individual wavefunction can be written as

\begin{equation}
\bra{\psi}M\ket{\psi}\label{eq:density matrix-1}
\end{equation}

where

\[
M_{i,j}=\begin{cases}
0 & \textrm{i or j =1}\\
{\displaystyle \sum_{i'}^{<\delta_{i,2},\delta_{i,3}>}}\ketbra{\phi_{i'}}{\phi_{i'}} & i=j\neq1\\
{\displaystyle \sum_{i'}^{<\delta_{i,2},\delta_{i,3}>}\sum_{j'}^{<\delta_{j,2},\delta_{j,3}>}}\ketbra{\phi_{i'}}{\phi_{j'}} & i\neq j
\end{cases},
\]

where $\delta_{i,j}$ is a Kronecker delta and <0,0> denotes the set
of states where neither site is occupied, <1,0> denotes the state
where site 1 is occupied and <0,1> denotes the state where site 2
is occupied. These single particle density matrices can then be combined
into a 4x4 total density matrix by taking the product of density matrix
elements to produce a combined density matrix using the following
rule

\begin{equation}
\rho_{i,j}^{comb.}={\displaystyle \sum_{p,q,r,s}}B_{i}^{pq}B_{j}^{rs}\rho_{pr}^{1}\rho_{qs}^{2}
\end{equation}

where B is a 4x4x4 tensor defined as

\begin{eqnarray*}
B_{1}^{pq} & = & \delta_{p=2q=3}+\delta_{p=3q=2}+\sum_{k=1}^{4}\delta_{p=1q=k}+\delta_{p=kq=1,}\\
B_{2}^{pq} & = & \delta_{p=2q=2}+\delta_{p=2q=4}+\delta_{p=4q=2,}\\
B_{3}^{pq} & = & \delta_{p=3q=3}+\delta_{p=3q=4}+\delta_{p=4q=3},\\
B_{4}^{pq} & = & \delta_{p=4q=4}.
\end{eqnarray*}

These rules can be applied repeatedly to build up a full multi-particle
density matrix. In this density matrix the fourth diagonal element
corresponds to no particles on either site, the third diagonal element
corresponds to a non-zero number of particles on site 2, but zero
particles on site 1, the second corresponds to none on site 2 but
a non-zero number on site 1, and the first corresponds to a non-zero
number on both sites.

\subsection{Estimate for long time mean of trace distance\label{sub:trace_mean}}

\begin{figure}
\includegraphics[scale=0.4]{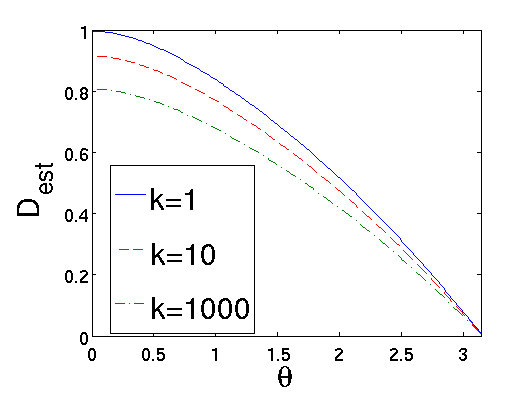}

\caption{Estimate of the long time mean of trace distance versus $\theta$,
the angle between $\psi_{0}$ and the x-axis with $\psi'_{0}$ chosen
as the orthogonal \emph{state} to $\psi_{0}$.\label{fig:angle_bound}}
\end{figure}

We now argue how the relevant non-Markovian dynamics of this system
can be illuminated by observing only a single pair of orthogonal states.
For the case of the spin system the choice of the two orthogonal initial
states is not as obvious as it was for the torus. As in the case of
2d geometries, every possible choice of initial state in of the system
has a single unique%
\footnote{Note that we want the initial \emph{states} to be orthogonal, not
the spin directions, i.e. the pair of \emph{states} in the +z and
-z directions are orthogonal, even though their spin vectors are actually
(anti-)parallel.%
} orthogonal partner. However, the best choice of initial state is
not clear a priori. In the following we explain the choice of initial
states for this setup. $H_{sys.}$ rotates spin vectors around the
x-axis. Therefore, a rotation of the initial pair of states about
this axis amounts to connecting the system to the bath at a different
time. Fig. \ref{fig:angle_bound} demonstrates that the average trace
distance should be smallest when $\psi_{0}$ is in the y-z plane.
We can understand this intuitively because initial states pointing
in the $\pm x$ direction are eigenstates of $H_{sys.}\otimes1_{bath}+1_{sys.}\otimes H_{bath}$,
whereas states in the y-z plane are only eigenstates of $1_{sys.}\otimes H_{bath}$.
We want to examine states in which the maximum information travels
into and out of the system, so we choose states in the y-z plane.

The estimate of the long time mean is calculated by totally dephasing
(i.e. setting all off diagonal elements in the energy basis to zero)
the total system-bath density matrix for the states $\psi_{0}$ and
$\psi'_{0}$ and calculating the trace distance between the resulting
system density matrices. Let us call the totally dephased matrix $\rho_{\infty}$,
for any operator A, $\textrm{Tr}[A\rho_{\infty}(\ket{\psi_{0}})]=\int_{\tau=0}^{\infty}d\tau\textrm{Tr}[A\exp(-iH\:\tau)\ketbra{\psi_{0}}{\psi_{0}}\{\exp(iH\:\tau))]$,
the long time mean of A. The trace distance can be though of as the
maximum possible distinguishibility between systems with the two initial
states using only measurements in the system, i.e. $D_{\psi\psi'}(t)=\max_{A}(\textrm{Tr}[A\textrm{Tr}_{bath}[\rho(t,\ket{\psi_{0}})]])-\textrm{Tr}[A\textrm{Tr}_{bath}[\rho(t,\ket{\psi'_{0}})]])$
where the eigenvalues of A are constrained to lie between 1 and 0.
We therefore have for the trace distance between $\textrm{Tr}_{bath}[\rho_{\infty}(\ket{\psi_{0}})]$
and $\textrm{Tr}_{bath}[\rho_{\infty}(\ket{\psi'_{0}})]$,

\begin{eqnarray*}
 &  & \max_{A}(\textrm{Tr}[A\textrm{Tr}_{bath}[\rho_{\infty}(\ket{\psi_{0}})]])-\textrm{Tr}[A\textrm{Tr}_{bath}[\rho_{\infty}(\ket{\psi'_{0}})]])\\
 & = & \max_{A}(\textrm{Tr}[A(\textrm{Tr}_{bath}[\rho_{\infty}(\ket{\psi_{0}})]-\textrm{Tr}_{bath}[\rho_{\infty}(\ket{\psi'_{0}})])]\\
 & = & \max_{A}(\textrm{Tr}[A(\textrm{Tr}_{bath}[\rho_{\infty}(\ket{\psi_{0}})-\rho_{\infty}(\ket{\psi'_{0}})]]\\
 & = & \max_{A}(\lim_{t\rightarrow\infty}\frac{1}{t}\int_{\tau=0}^{t}d\tau(\textrm{Tr}[A(\exp(-iH\:\tau)\ketbra{\psi_{0}}{\psi_{0}}\exp(iH\:\tau))\\
 & - & \exp(-iH\:\tau)\ketbra{\psi'_{0}}{\psi'_{0}}\exp(\imath H\:\tau))]))\\
 & \lessapprox & \lim_{t\rightarrow\infty}\frac{1}{t}\int_{\tau=0}^{t}d\tau(\max_{A}(\textrm{Tr}[A(\exp(-iH\:\tau)\ketbra{\psi_{0}}{\psi_{0}}\exp(iH\:\tau))\\
 & - & \exp(-iH\:\tau)\ketbra{\psi'_{0}}{\psi'_{0}}\exp(iH\:\tau))]))\\
 & = & \lim_{t\rightarrow\infty}\frac{1}{t}\int_{\tau=0}^{t}d\tau D_{\psi\psi'}(\tau).
\end{eqnarray*}

The trace distance between $\textrm{Tr}_{bath}[\rho_{\infty}(\ket{\psi_{0}})]$
and $\textrm{Tr}_{bath}[\rho_{\infty}(\ket{\psi_{0}'})]$ gives the
maximum average distinguishability between $\rho_{s}(t,\ket{\psi_{0}})$
and $\rho_{s}(t,\ket{\psi_{0}'})$ assuming measurements are performed
in the same basis at all times. The trace distance between the totally
dephased states provides a lower bound for, and estimate of the average
trace distance between $\rho_{s}(t,\ket{\psi_{0}})$ and $\rho_{s}(t,\ket{\psi_{0}'})$
which is the maximum average distinguishability if measurements can
be made in a different basis at every time.

\subsection{Scaling arguments for $t_{g}$versus k\label{sub:t_g-k}}

We can make very similar arguments to those given in Sec. \ref{sec:spect_args}
of the main paper for why $t_{g}\propto\frac{1}{k^{2}}$. First let
us consider the Hamiltonian of a system coupled to a bath in the canonical
form

\begin{equation}
H=H_{sys.}\otimes1_{bath}+1_{sys.}\otimes H_{bath}+k\: H_{sys.}^{coup}\otimes H_{bath}^{coup}.\label{eq:canon_form}
\end{equation}
 Note that Hamiltonian given in Eq. \ref{eq:conn Ham} is of this
form. We now note that we can decompose $H_{sys.}^{coup}\otimes H_{bath}^{coup}=D_{sys.}^{coup}\otimes D_{bath}^{coup}+R_{sys.}^{coup}\otimes R_{bath}^{coup}$
where $[D_{sys.}^{coup},H_{sys.}]=0$ and $[D_{bath}^{coup},H_{bath}]=0$.
We now note that we can write $H_{sys.}\otimes1_{bath}+1_{sys.}\otimes H_{bath}+k\:(D_{sys.}^{coup}\otimes D_{bath}^{coup}+R_{sys.}^{coup}\otimes R_{bath}^{coup})=H_{1}(k)+k\: R_{sys.}^{coup}\otimes R_{bath}^{coup}$
where $H_{1}(k)$ cannot equilibrate the bath by itself because $[D_{sys.}^{coup}\otimes D_{bath}^{coup},H_{sys.}\otimes1_{bath}]=0$
and $[D_{sys.}^{coup}\otimes D_{bath}^{coup},1_{sys.}\otimes H_{bath}]=0$.
In other words, a gap degeneracy will persist in $H_{1}(k)$ for any
value of k. The gap degeneracy is broken by the term $k\: R_{sys.}^{coup}\otimes R_{bath}^{coup}$,
however $R_{sys.}^{coup}\otimes R_{bath}^{coup}$ can always be chosen
such that it contains only off diagonal elements in the energy basis
of $H_{1}(k=0)$. The off diagonal elements can only change the spectrum
though level repulsion rather than direct level splitting. The deviation
from gap degeneracy will therefore scale like $k^{2}$rather than
k, and as a result we observe $t_{g}\propto\frac{1}{k^{2}}$.

\subsection{Eigenstates and eigenvalues of fully connected graph\label{sub:full_spect}}

Consider the Hamiltonian of a fully connected tight binding graph
with equal coupling between all sites and zero on-site potential,
$H_{c}=\sum_{i,j}^{N}c_{i}^{\dagger}c_{j}$. This Hamiltonian has
a N-1 fold degenerate ground state manifold that spans all states
for which the total amplitude sums to zero, i.e. $\sum_{i}\psi_{i}=0$.
This can be shown in the case of zero on-site potential by observing
that $(H_{c}\psi)_{i}=\sum_{j\neq i}\psi_{j}$. Therefore in the case
where $\sum_{i}\psi_{i}=0$, we see that $(H_{c}\psi)_{i}=\sum_{j}\psi_{j}-\psi_{i}=-\psi_{i},$
these states therefore are eigenstates with eigenvalue -1. In the
case where $\psi_{i}=\psi_{j}=\frac{\exp(\imath\phi)}{\sqrt{N}}$
we have $(H_{c}\psi)_{i}=\sum_{j\neq i}\psi_{j}=(N-1)\psi_{i}$, this
state is therefore an eigenstate with an eigenvalue of N-1.

Now let us now consider a Hamiltonian of the form $H_{comb.}=H_{c}\oplus H_{a}+H_{join}$,
where $H_{join}=\sum_{i=1}^{N}\sum_{j=N+1}^{N+M}c_{i}^{\dagger}c_{j}$
equally couples all N sites in $H_{c}$ to another group of sites,
$H_{a}$ is an arbitrary Hamiltonian of size M. We now note that for
an initial state $\psi$$^{trap}$ such that $\sum_{i=1}^{N}\psi_{i}^{trap}=0$
and $\psi_{i}=0\;\forall\: i>N$. We can show that $\psi$$^{trap}$
is an eigenstate of $H_{c}\oplus H_{a}$ because it has no support
for $i>N$, and the part for $i\leq N$ is an eigenstate of $H_{c}$
with an eigenvalue of (N-1). We now also see that $(H_{join}\psi^{trap})_{i\leq N}=0$
and $(H_{join}\psi^{trap})_{i>N}=\sum_{i=1}^{N}\psi_{i}^{trap}=0$,
therefore $H_{comb}$ will have an N-1 fold degenerate manifold of
states which only have finite amplitudes in the first N sites and
will have an energy of N-1. Wavefunctions in these states are effectively
trapped in the first N sites.

We know by orthogonality of eigenstates, that all other eigenstates
of $H_{comb}$ must have $\psi_{i}=\psi_{j}=\psi_{cc}$ for all i
and j less then N+1. This constraint means that the internal degrees
of freedom of the first N sites do not play a role in the eigenstates
outside of the manifold mentioned in the previous paragraph. The first
N sites can therefore be modeled as a single site.


\begin{thebibliography}{10}
\bibitem{Breuer(2002)}H. P. Breuer and F. Petruccione, The Theory
of Open Quantum Systems (Oxford University Press, 2002)

\bibitem{Rivas(2012)}Á. Rivas and S. F. Huelga, Open Quantum Systems
An Introduction (Springer, Berlin Heidelberg, 2012)

\bibitem{Krovi(2007)}H.Krovi, O. Oreshkov, M. Ryazanov, and D.A.
Lidar, Phys. Rev. A 76,052117 (2007).

\bibitem{Boccheri(1957)}P. Bocchieri and A. Loinger, Phys. Rev. 107,
337 (1957)

\bibitem{Venuti2010}L.C. Venuti and P. Zanardi, Phys. Rev. A 81,
032113 (2010)

\bibitem{Wolf(2008)}M. M. Wolf, J. Eisert, T. S. Cubitt and J. I.
Cirac, Phys.Rev. Lett. 101, 150402 (2008).

\bibitem{Shabani(2009)}A. Shabani and D. A. Lidar, Phys. Rev. Lett.
102, 100402 (2009).

\bibitem{Wilkie(2009)}J. Wilkie and Y. M. Wong, J. Phys. A 42, 015006
(2009).

\bibitem{Budnini(2006)}A. A. Budini, Phys. Rev. A 74, 053815 (2006)

\bibitem{Rivas(2010)}Á. Rivas, S. F. Huelga, and M. B. Plenio, Phys.
Rev. Lett. 105, 050403 (2010)

\bibitem{Breuer(2004)}H. P. Breuer, Phys. Rev. A 70, 012106 (2004).

\bibitem{Breuer2009}H.P. Breuer, E.M. Laine, J Piilo PRL 103, 2104021
(2009)

\bibitem{Laine(2010)}E. M. Laine, J. Piilo and H. P. Breuer, Phys.
Rev. A 81, p. 062115 (2010)

\bibitem{Breuer(2012)}H. P. Breuer, J. Phys. B: At. Mol. Opt. Phys.
45 (2012)

\bibitem{Wissmann2012}S. Wissmann, A.Karlsson, E.M. Laine, J. Piilo,
H.P. Breuer arXiv:1209.4989 (2012)

\bibitem{Piilo(2008)}J. Piilo, S. Maniscalco, K. Härk\"{ö}nen and
K.-A. Suominen, Phys. Rev. Lett. 100, 180402 (2008); J. Piilo, K.
Härk\"{ö}nen, S. Maniscalco and K.-A. Suominen, Phys. Rev. A 79,
062112 (2009); H. P. Breuer and J. Piilo, Europhys. Lett. 85, 50004
(2009)

\bibitem{Breuer(2009-1)}H. P. Breuer and B. Vacchini, Phys. Rev.
Lett. 101, 140402 (2008); Phys. Rev. E 79, 041147 (2009).

\bibitem{vanWonderen(2006)}A. J. van Wonderen and K. Lendi, Europhys.
Lett. 71, 737 (2006).

\bibitem{Barnett(2001)}S. M. Barnett and S. Stenholm, Phys. Rev.
A 64,033808 (2001).

\bibitem{Kossakowski(2009)}A. Kossakowski and R. Rebolledo, Open
Syst. Inf. Dyn. 15, 135 (2008); Open Syst. Inf. Dyn. 16, 259 (2009).

\bibitem{Churscinski(2010)}D. Chruściński, A. Kossakowski and S.
Pascazio, Phys. Rev. A 81, 032101 (2010).

\bibitem{Daffler(2010)}S. Daffer, K. Wódkiewicz, J. D. Cresser and
J. K. McIver, o Phys. Rev. A 70, 010304(R) (2004).

\bibitem{Mata(2012)} I. Garca-Mata, C. Pineda and D. Wisniacki, Phys.
Rev. A 86, 022114 (2012).

\bibitem{Chiuri(2012)}A. Chiuri, C. Greganti, L. Mazzola, M. Paternostro
and P. Mataloni, Scientific Reports 2, 968 (2012)

\bibitem{Liu(2011)}B.-H. Liu, L. Li, Y.-F. Huang, C.-F. Li, G.-C.
Guo, E.-M. Laine, H.-P. Breuer and J. Piilo, Nature Physics 7, 931934
(2011).

\bibitem{Ubbelohde(2012)} N. Ubbelohde, K. Roszak, F. Hohls, N. Maire,
R. J. Haug \& T. Novotný, Scientific Reports 2, 374 (2012)

\bibitem{Huelga(2012)} S. F. Huelga, A. Rivas and M. B. Plenio, Phys.
Rev. Lett. 108, 160402 (2012).

\bibitem{Alonso(2005)}D. Alonso and I. de Vega, Phys. Rev. Lett.
94, 200403 (2005).

\bibitem{Diosi(2012)}L. Diósi, Phys. Rev. A, 85, 034101 (2012).

\bibitem{Gardiner(1985)}C. W. Gardiner and M. J. Collett, Phys. Rev.
A 31, 3761–3774 (1985).

\bibitem{Jack(1999)}M. W. Jack, M. J. Collett, and D. F. Walls, J
opt. B 1, 452 (1999).

\bibitem{Jack(2000)}M.W.Jack and M.J.Collett, Phys. Rev. A 61, 062106
(2000).

\bibitem{Diosi(2008)}L. Diósi, Phys. Rev. Lett. 100, 080401 (2008).

\bibitem{Wiseman(2008)}H. M. Wiseman and J. M. Gambetta, Phys. Rev.
Lett. 101, 140401 (2008).

\bibitem{Diosi(2008)-1}L. Diósi, Phys. Rev. Lett. 101, 149902(E)
(2008).

\bibitem{Mazzola(2009)}L. Mazzola, S. Maniscalco, J. Piilo, K.-A.
Suominen, and B. M. Garraway, Phys. Rev. A 80, 012104 (2009).

\bibitem{Gambetta(2003)}J. Gambetta and H. M. Wiseman, Phys. Rev.
A 68, 062104 (2003).

\bibitem{Florescu(2001)}M. Florescu and S. John, Phys. Rev. A. 64,
33801 (2001).

\bibitem{John(1995)}S. John, T. Quang, Phys. Rev. Lett. 74, 3419
(1995).

\bibitem{de Vega(2008)}I. de Vega et. al PRL 101, 260404 (2008).

\bibitem{Diez (2010)}M. Diez, N. Chancellor, S. Haas, L.C. Venuti,
P. Zanardi, Phys. Rev. A 82, 032113 (2010)

\bibitem{vonNeumann(1929)}J. von Neumann, Zeitschrift für Physik
57, 30 (1929), see also the english translation by R. Tumulka at arXiv:1003.2133.

\bibitem{Tasaki(1998)}H. Tasaki, Phys. Rev. Lett. 80, 1373 (1998).

\bibitem{Popescu(2006)}S. Popescu, A. J. Short, and A. Winter, Nature
Physics 2, 754 (2006).

\bibitem{Linden(2009)}N. Linden, S. Popescu, A. J. Short, and A.
Winter, Phys. Rev. E 79, 061103 (2009).

\bibitem{Reimann(2008)}P. Reimann, Phys. Rev. Lett. 101, 190403 (2008).

\bibitem{Goldstein(2010)}S. Goldstein, J. L. Lebowitz, R. Tumulka,
, and N. Zanghí (2010), 1003.2129.

\bibitem{Nielsen and Chuang} M. A. Nielsen and I. L. Chuang, Quantum
Computation and Quantum Information (Cambridge University Press, Cambridge,
2000)

\bibitem{Kosmann-Schwarzbach (2010}Kosmann-Schwarzbach, Yvette. From
Finite Groups to Lie Groups (Springer, New York 2010)\end{thebibliography}
\end{document}